\RequirePackage{fixltx2e}
\documentclass[twocolumn,secnumarabic,amssymb, superscriptaddress, nobibnotes, aps, prd]{revtex4-1}

\usepackage[T1]{fontenc}
\usepackage[utf8]{inputenc}
\usepackage[english]{babel}
\usepackage{lmodern}
\usepackage{amsmath}
\usepackage{amsfonts}
\usepackage{amssymb}
\usepackage{graphicx}
\usepackage{fixltx2e}
\usepackage{caption}
\usepackage{subcaption}
\usepackage[usenames,dvipsnames,svgnames,table]{xcolor}
\usepackage{etoolbox}
\usepackage{blindtext}
\usepackage{gensymb}
\usepackage{textcomp}
\usepackage{enumitem}
\usepackage{float}
\usepackage{dsfont}
\usepackage{cases}
\usepackage{url}
\usepackage{textcase}
\usepackage{bm}
\usepackage{epstopdf}
\newcommand{\eqv}[1]{\begin{equation} #1 \veq \end{equation}}
\newcommand{\eqp}[1]{\begin{equation} #1 \peq \end{equation}}

\newcommand{\und}[1]{\underline{ #1 }}
\newcommand{\undd}[1]{\underline{\underline{ #1 }}}
\newcommand{\dd}[2]{\dfrac{\mathrm d \,#1}{\mathrm d\,  #2}}

\newcommand{\un}[1]{\,\mathrm{#1}}

\newcommand{\ind}[1]{_{\mathrm{#1}}}

\newcommand{\fig}[1]{Figure~\ref{#1}}

\newcommand{\mum}{\, \mathrm{\mu m}}
\newcommand{\mnm}{\, \mathrm{mN/m}}
\newcommand{\etal}{\textit{et al.}}
\newcommand{\pd}[1]{\cdot 10^{#1}}

\newcommand{\peq}{\; \; .}
\newcommand{\veq}{\; \; ,}
\DeclareGraphicsExtensions{.eps,.pdf,.png,.jpg}

\makeatletter
\newcommand{\mypm}{\mathbin{\mathpalette\@mypm\relax}}
\newcommand{\@mypm}[2]{\ooalign{%
		\raisebox{.1\height}{$#1+$}\cr
		\smash{\raisebox{-.6\height}{$#1-$}}\cr}}
\makeatother

\begin{document}
\title{Transient deformation of a droplet near a microfluidic constriction : a quantitative analysis}

\author{Corentin Tr\'egou\"et}
\affiliation{UMR CNRS SIMM 7615, ESPCI Paris, PSL Research University, 75005 Paris, France.}
\affiliation{UMR CNRS Gulliver 7083, ESPCI Paris, PSL Research University, 75005 Paris, France.}
\author{Thomas Salez}
\email[corresponding author:]{thomas.salez@u-bordeaux.fr}
\affiliation{UMR CNRS Gulliver 7083, ESPCI Paris, PSL Research University, 75005 Paris, France.}
\affiliation{Univ. Bordeaux, CNRS, LOMA, UMR 5798, F-33405 Talence, France.}
\affiliation{Global Station for Soft Matter, Global Institution for Collaborative Research and Education, Hokkaido University, Sapporo, Japan.}
\author{C\'ecile Monteux }
\affiliation{UMR CNRS SIMM 7615, ESPCI Paris, PSL Research University, 75005 Paris, France.}
\affiliation{Global Station for Soft Matter, Global Institution for Collaborative Research and Education, Hokkaido University, Sapporo, Japan.}
\author{Mathilde Reyssat}
\email[corresponding author:]{mathilde.reyssat@espci.fr}
\affiliation{UMR CNRS Gulliver 7083, ESPCI Paris, PSL Research University, 75005 Paris, France.}

\begin{abstract}
We report on experiments that consist in deforming a collection of monodisperse droplets produced by a microfluidic chip through a flow-focusing device. We show that a proper numerical modelling of the flow is necessary to access the stress applied by the latter on the droplet along its trajectory through the chip. This crucial step enables the full integration of the differential equation governing the dynamical deformation, and consequently the robust measurement of the interfacial tension by fitting the experiments with the calculated deformation. Our study thus demonstrates the feasibility of quantitative in-situ rheology in microfluidic flows involving \textit{e.g.} droplets, capsules or cells.
\end{abstract}
\date{\today}
\maketitle

\section{Introduction}
Droplet-based microfluidics enables the fragmentation of compounds (chemicals, cells…) into nanoliters. Such micro-droplets can be further protected by a thin elastic shell realized through several micro-encapsulation techniques. The inner phase of such capsules can then be easily transported and delivered into specific places \cite{Shum2008,Seemann2012,Parker2015,DoNascimento2016,Wang2017a}. At the microscale, the deformation of droplets or capsules is controlled by a balance between the driving forces which deform the objects (shear, elongation…) and the restoring ones results from their interfacial properties: the interfacial tension in the case of droplets or the interfacial elasticity in the case of capsules, as well as the bulk and interfacial viscosities. Measuring the interfacial properties \textit{in situ} in a microfluidic channel is particularly relevant as it can give insights into the dynamical composition, stability, and rheology of such interfaces (including encapsulation membranes), which is crucial to understand or optimize their behaviour. 

For simple droplets, different approaches have been investigated in the past to measure interfacial tension by shear deformation. The first quantitative experiments of droplet deformation under viscous stress were performed by Taylor in 1934~\cite{Taylor1934}, using the Four-roller apparatus. Therein, the fluid arrives in a 2D cross-shaped location from two opposite sides and leaves on the two orthogonal others. The droplet is thus sheared and deformed. The steady shape in that case results from a balance between the driving viscous stress and the restoring interfacial tension. Consequently, the knowledge of the viscous shear stress at the interface and the measurement of the deformation lead to a measurement of the interfacial tension.

Lee \textit{et al.}~\cite{Lee2007} and Deschamps \textit{et al.}~\cite{Deschamps2009} miniaturized Taylor's experiment on a microfluidic chip to measure interfacial properties of microdroplets and vesicles. More recently, Xie \textit{et al.}~\cite{Xie-de-Loubens} use similar geometry to characterize the interfacial rheological properties of microcapsules. Hudson \textit{et al.} further integrated this device into a microfluidic channel by using a convergent and then divergent channel profile~\cite{Hudson2005,Cabral2006}. That device generates a gradient of velocity in the main direction of the flow, leading to shear stress and thus droplet deformation. This in-flow process then allows to perform measurements on a large number of droplets, and hence enables a statistical treatment of the data which enhances the accuracy of the results. The previous setup has been further adapted by Brosseau \textit{et al.}~\cite{Brosseau2014} in order to deform capsules, which require a stronger stress. This last geometry is composed of an alternating succession of narrow channels, in which the capsules are confined by the walls, and wide chambers where they only interact with the viscous flow. The transition between two parts is sharp in the latter geometry.

Hudson \etal~have shown that the droplets in the in-flow device are in a transient regime \cite{Hudson2005, Cabral2006}. In specific conditions (geometry, viscosity contrast), the authors have succeeded in extracting the interfacial tension by dynamically analyzing the shape of drops but their approach relies on a derivative of the experimental data -- which can induce large errors. However, the evolution of such non-stationary shapes has been described theoretically by Barthes-Biesel \etal~\cite{BarthesBiesel1973} which could in principle be used to get interfacial tension in any configuration. But up to now, the precise evolution of the shape of a droplet in any geometry, and its fitting to theoretical predictions has not been performed. 

In the present study, we perform experiments that consist in deforming a collection of monodisperse droplets produced by a microfluidic chip through a flow-focusing device. We show that a proper numerical modelling of the flow is necessary to access the stress applied by the latter on the droplet along its trajectory through the chip. This crucial step enables the full integration of the differential equation governing the dynamical deformation, and consequently the robust measurement of the interfacial tension by fitting the experiments with the calculated deformation.

\section{Material and methods}
\subsection{Materials}
The droplets consist of mineral oil (Sigma Aldrich,  Mineral Oil Rotational Viscosity Standard of viscosity $29.04 \un{mPa.s}$ at $25.00 \un{\degree C}$) flowing in a polymer solution (poly(methacrylic acid), PMAA, provided by Polysciences) at $1 \un{\% w}$ adjusted at pH 3. The droplets are characterized in the same polymer solution (PMAA $1 \un{\% w}$, pH 3). They are produced in microfluidics using a standard flow-focusing device made of poly(dimethyl siloxane) (PDMS). Such a setup provides a very monodisperse collection of droplets ranging from 50 to 70 $\mum$ in diameter. The characterization chips are realized with a photosensitive adhesive provided by Norland (NOA 81). This microfabrication technique allows to build non-deformable channels confined between two glass surfaces, and able to sustain large flow rates without being deformed~\cite{Bartolo2008}. The inlet is connected to the droplet-production chip by a silicon tubing (Tygon) of inner diameter $800 \mum$, coated with a solution of Bovine Serum Albumine (BSA) provided by Sigma Aldrich, to prevent adhesion of the droplets inside the tubing (incubation at $1 \un{\%w}$ solution during $12 \un{h}$ at room temperature). The outlet is connected to a Peek tubing of inner diameter $125 \mum$. 

\subsection{Methods}
\subsubsection{Observation of the deformation} \label{geometry}
\begin{figure}[t!]
	\begin{subfigure}[b!]{0.5 \textwidth}
		\includegraphics[width=1\textwidth]{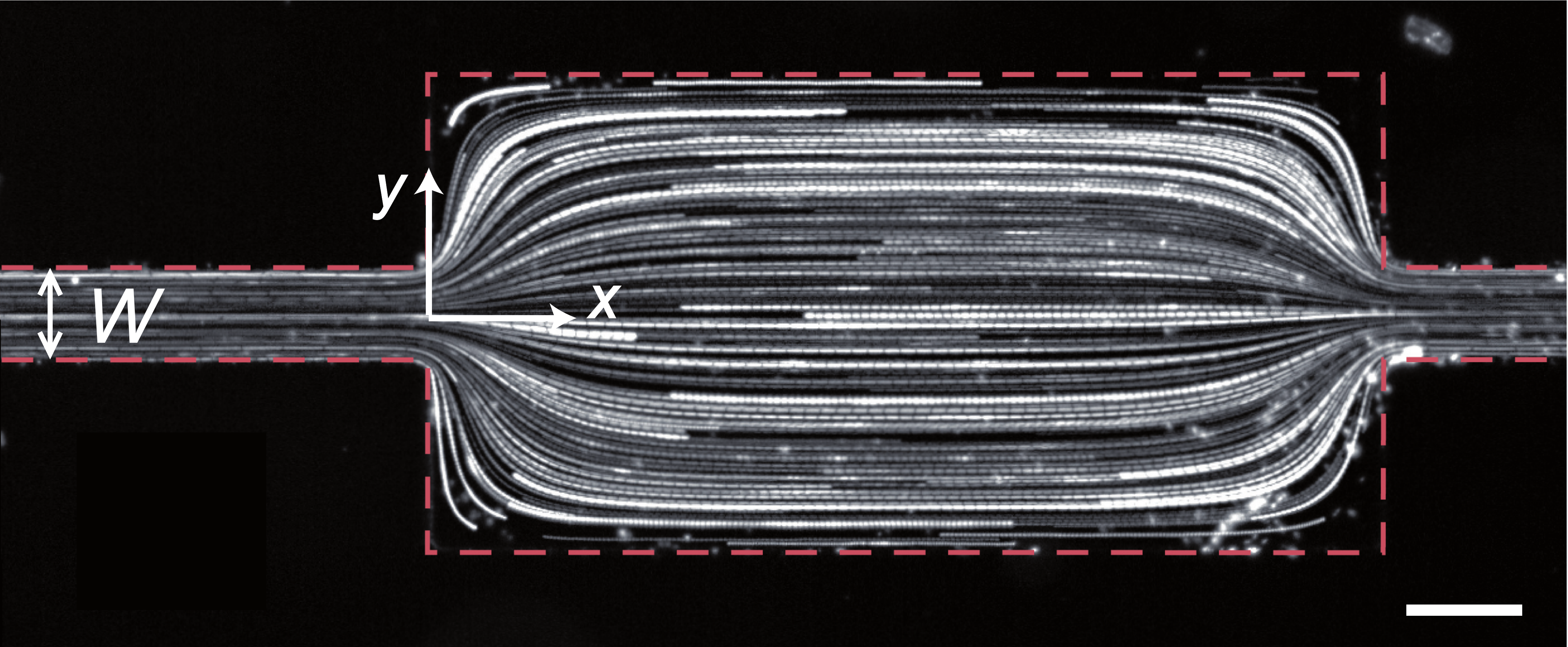}
		\caption{\label{figure1a}}
	\end{subfigure} 
	\\
	\begin{subfigure}[b!]{0.3 \textwidth}
		\includegraphics[height=2.8 cm]{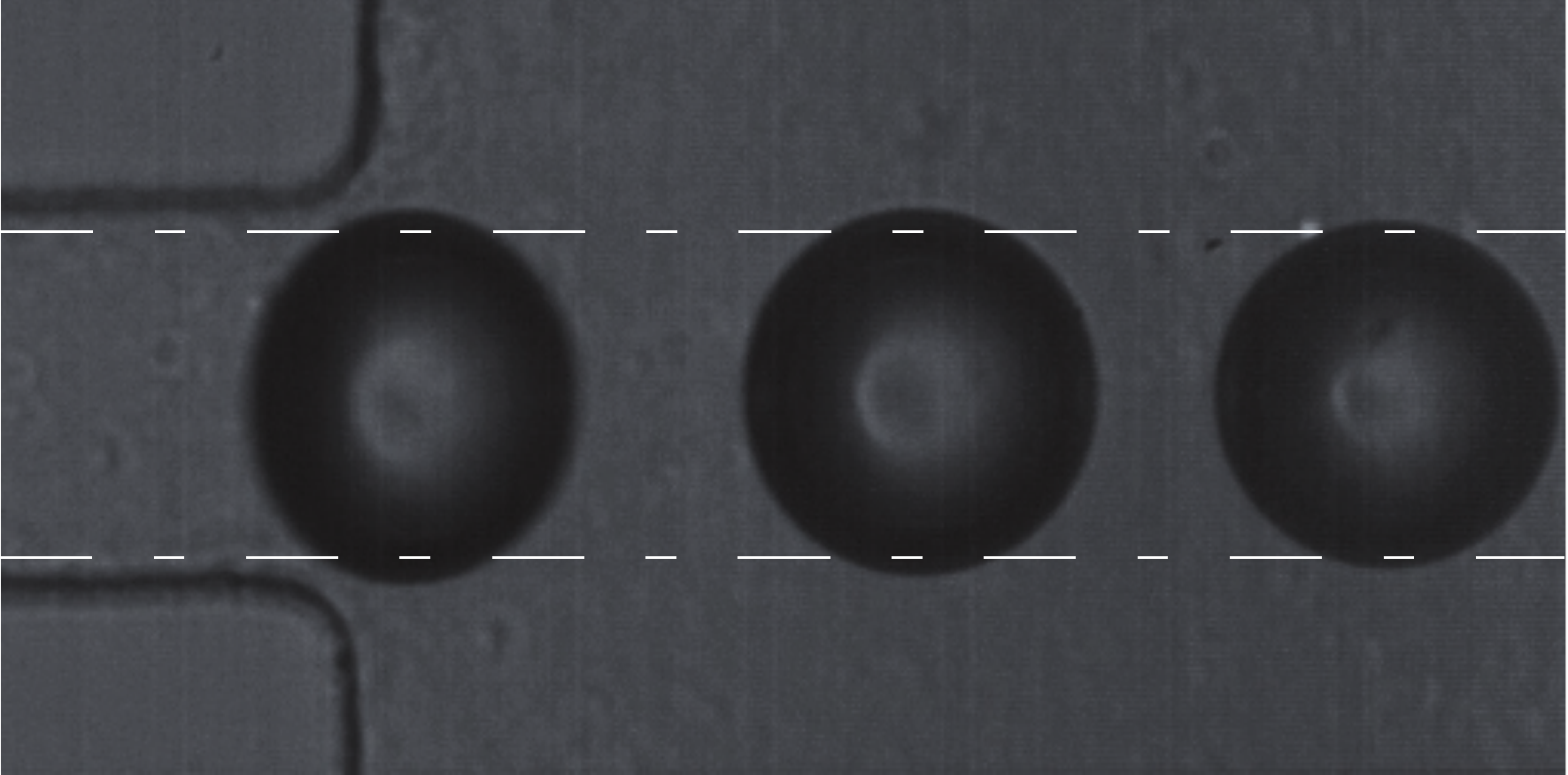}
		\caption{\label{figure1b}}
	\end{subfigure}
	\hfill
	\begin{subfigure}[b!]{0.17 \textwidth}
		\includegraphics[height=2.8 cm]{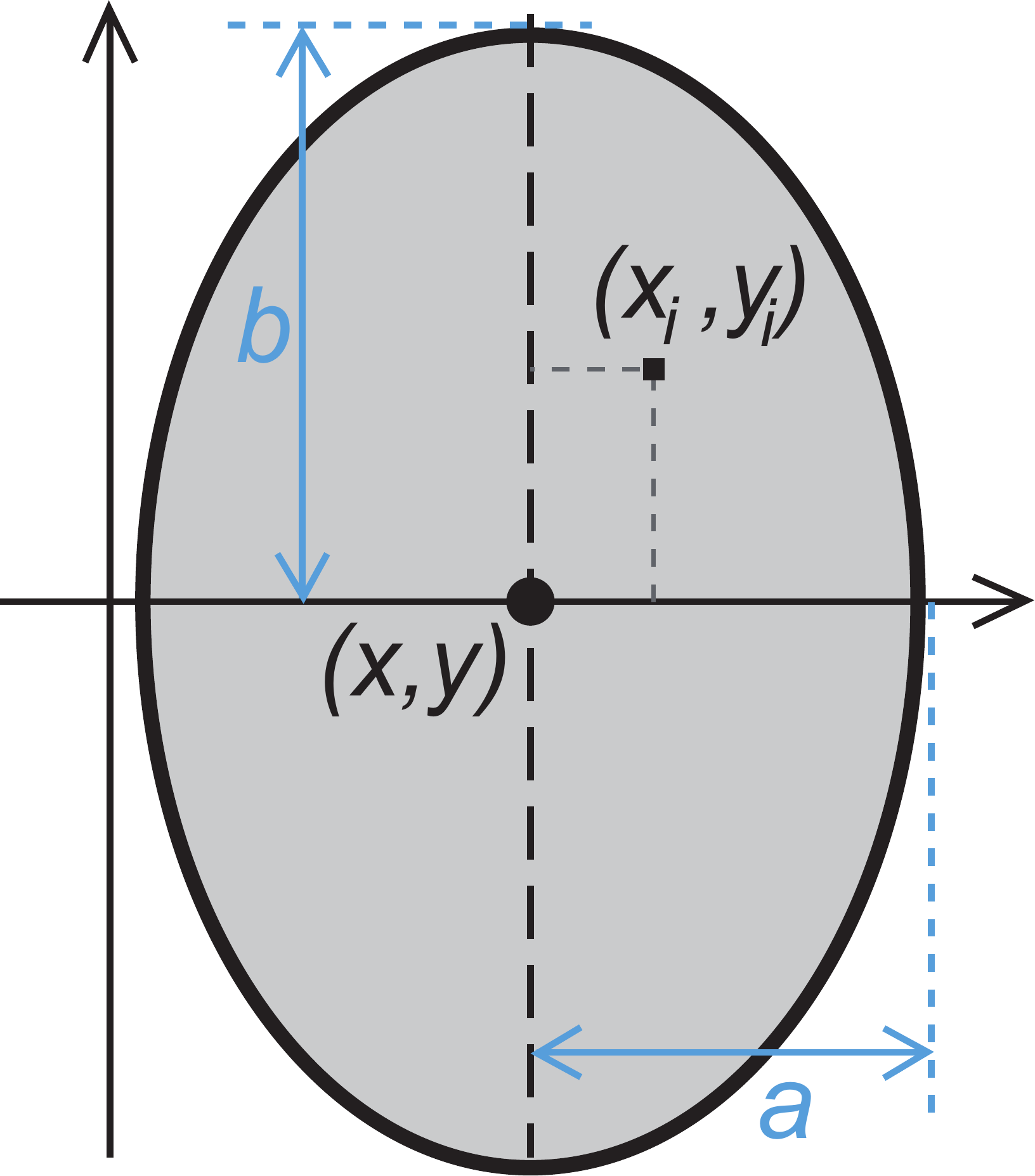}
		\caption{\label{figure1c}}
	\end{subfigure}
\caption{\label{figure1}(a) Characterization chip. Streamlines (from left to right) are visible due to fluorescent tracers and show the divergent flow at the entrance of the wide chamber. Scale bar is $100 \mum$. (b) Droplet entering the wide chamber. The droplet is first elongated along $y$ and then relaxes towards a spherical shape (stack of 3 images). The deformation of a few \% is barely visible by eye.  (c) Schematics showing the parameters $a$ and $b$ describing the droplet shape.}
\end{figure}

The characterization chip consists in a sharp transition between a narrow channel (of width $W=40 \mum$) and a wide chamber (of width $3 W$), similar to the chip used by Polenz \textit{et al.}~\cite{Polenz2015}. Such a geometry imposes a divergent flow near the entrance of the wide chamber, as illustrated by fluorescent tracers in \fig{figure1a}. The height in the $z$-direction is constant and equals $100\un{\mu m}$ which prevents the droplets from being confined in the $z$-direction.

The divergent flow at the entrance of the wide chamber generates a viscous stress on the droplets, which tends to elongate them perpendicularly to the flow direction. We note that the convergent flow at the exit of the wide chamber could also be used to deform the droplets. However, the narrow channel better aligns the droplets with the center of the wide chamber at its entrance than at its exit -- especially when the droplets are laterally confined in the narrow channel, which is the case in our experiments. As a consequence, for more reproducible results, observations and measurements are performed only at the entrance of the wide chamber. The flow is controlled by a pressure controller Fluigent.

Observation is performed along the $z$-direction through a Leica inverted microscope with a 10$\times$ objective. A Photron Fastcam-SA camera mounted on the microscope allows to record the droplet deformation at high acquisition frequency (typically  $10\,000 \un{fps}$). Pictures similar to the ones presented in~\fig{figure1b} are recorded every $0.1 \un{ms}$ during $2 \un{s}$. 
The different steps of the image processing needed to get the deformation measurement as a function of position in the wide chamber are detailed in the Appendix. 

\subsubsection{Governing equation}
\label{TheoryHudson}
To relate the measured deformation to interfacial tension, Taylor developed a theoretical model~\cite{Taylor1934}, which was generalized by Barthes-Biesel \etal~\cite{BarthesBiesel1973,BarthesBiesel1981}, and summarized by Rallison~\cite{Rallison1984}. This model has been used by Hudson \etal \cite{Hudson2005} in order to extract interfacial tension values.

The model of Taylor \cite{Taylor1934} considers a droplet of viscosity $\eta_{\mathrm{droplet}}$ placed in a fluid medium of viscosity $\eta_{\mathrm{medium}}$. The radius of the droplet at rest is $r$, and the interfacial tension with the surrounding fluid is $\gamma$. In Taylor's model, the latter fluid undergoes an hyperbolic flow in the $(x,y)$-plane with a velocity field that reads $\und{u}=(u_x,u_y)=C\cdot (x,-y)$, where $C$ is a constant fixed by the flux. The extension rate $\dot \epsilon$ along $x$ is thus related to the gradient of velocity of the fluid along $x$, through: $\dot \epsilon= \frac{\textrm{d}u_x}{\textrm{d}x}=C$. In this configuration, $\dot \epsilon$ does not depend on the position, and the steady deformation $D_{\mathrm{{steady}_{Taylor}}}$ (see definition of the deformation $D$ in Appendix) of the droplet is directly proportional to $ \dot \epsilon$~\cite{Taylor1934}, as: 
\begin{equation}D_{\mathrm{{steady}_{Taylor}}}=\dfrac{19 \lambda +16}{16 \lambda +16} \cdot \dfrac{\eta_{\mathrm{medium}} \, r}{\gamma}\cdot \dot \epsilon \label{Dsteady-Taylor}\; , \end{equation}
where $\lambda=\frac{\eta_{\mathrm{droplet}}}{\eta_{\mathrm{medium}}}$ is the viscosity ratio between the droplet and the fluid medium. 

When the flow is less regular than the hyperbolic flow described by Taylor, Barthes-Biesel \etal~\cite{BarthesBiesel1973} state that for moderated shear stress and hence deformation, the flow in the vicinity of the droplet can be approximated to the first order in deformation and described by a deformation-rate tensor independent of position. In these conditions the steady deformation must be calculated with the eigenvalues $e_{1}$ and $e_{2}$ of the deformation-rate tensor (see definition in Equation~\eqref{BidimensionnalDeformationRate}), through: 
\begin{equation}D_{\mathrm{{steady}}}=\dfrac{19 \lambda +16}{16 \lambda +16} \cdot \dfrac{\eta_{\mathrm{medium}} \, r}{\gamma}\cdot  (e_{1}-e_{2})\label{Dsteady}\; . \end{equation}
Furthermore, Barthes-Biesel \textit{et al.}~\cite{BarthesBiesel1973} describe the transient regime as a first-order relaxation towards the steady state of Equation~\eqref{Dsteady}. In this framework, the deformation $D(t)$ satisfies:
\begin{equation}\dfrac{\mathrm d D}{\mathrm d t}=\dfrac{1}{\tau_{\mathrm{ca}}}\cdot \left(D_{\mathrm{steady}}-D\right)\label{TransientD}\; ,\end{equation}
with a relaxation time $\tau_{\mathrm{ca}}$ defined as:
\begin{equation}\tau_{\mathrm{ca}}=\dfrac{2}{5}\cdot (2 \lambda+3) \cdot\dfrac{19 \lambda +16}{16 \lambda +16} \cdot \dfrac{\eta_{\mathrm{medium}} \, r}{\gamma}\label{LiquidRelaxationTimeCapsule}\; . \end{equation}
The droplet-medium interfacial tension $\gamma$ can thus be obtained by fitting the experimental deformation $D(t)$ of the droplet using Equations~(\ref{TransientD}) and~(\ref{LiquidRelaxationTimeCapsule}). The only requirement is to know precisely the position-dependent steady deformation $D\ind{steady}$ of the droplet in the considered geometry. 

\subsubsection{Calculation of the steady deformation}
\label{simu}
Our geometry is described in \ref{geometry}. In absence of any symmetry, and including further a moving droplet, it is not straightforward to know precisely the viscous stress exerted by the surrounding fluid on the droplet. Performing finite-element numerical simulations is a robust way to access this information, in order to further calculate the position-dependent steady deformation $D\ind{steady}$ of the droplets. The flow field and the associated velocity gradient are extracted from such simulations (using the software Comsol Multiphysics$^{\mbox{\scriptsize{\textregistered}}}$), and analyzed to calculate the steady deformation of the droplets.

\paragraph*{Framework of the simulations.}
The simulated geometry consists of a fixed hard sphere placed at the position $(x\ind{0},0,0)$ near the entrance of a wide chamber three times larger (in $y$) than the incoming narrow channel, but with identical height (in $z$). The hard-sphere idealization is justified by the tiny experimental deformations that alter only marginally the flow. Calibration experiments are performed with tracers (of diameter $5\mum$) in order to know precisely the relative speed between the medium and the droplet. It appears that at a distance $x=150 \mum$, the droplet is nearly spherical and its velocity is similar to the medium speed. Consequently, even without tracers, the rest diameter of the droplet can be measured in the far field, while the far-field speed of the droplet indicates the flow velocity and thus the flow rate. Therefore, in every simulation, the droplet diameter and the flow rate can be set to mimic precisely a given experiment.
To analyze a given experiment, simulations are made for different positions of the droplet (typically $10$ points between $x=20\mum$ and $x=120\mum$). In such simulations, the droplet does not move by design, as we only consider steady states in this part of the modelling.

\paragraph*{Boundary conditions.}
We assume a no-slip condition at the walls, which implies $\und{u}=\und{0}$ there since the walls are not permeable/deformable. That assumption is valid, since the typical length scales of our microfluidic device are much larger than the slip length in a water-like solution near a glassy wall, as the latter slip length seldom exceeds $100\un{nm}$, according to Lauga \textit{et al.}~\cite{Lauga2005}. In contrast, the boundary condition at the droplet surface is not straightforward. First, we assume that there is no slip between the medium and the droplet, and that there is no velocity in the droplet fluid at the droplet surface. These assumptions are motivated by our high viscosity ratio $\lambda=30$, together with the slow dynamics of the polymers at the interface. 
Eggleton et al \cite{eggleton_pawar_stebe_1999} showed that a no-slip boundary condition at the interface is obtained when the viscous drag which tends to cause polymer concentration gradients is compensated by a Marangoni flow in the opposite direction resulting in a zero tangential velocity. According to \cite{eggleton_pawar_stebe_1999}, even very small surface tension gradients can compensate viscous drag. Moreover, the Peclet number which represents the ratio of convective motion over diffusive motion is very high in our experiment, $\textrm{Pe}=\frac{||\und{u}|| \cdot R}{D\ind{diffusion}}\sim 10^5 $ where $D\ind{diffusion}\sim 10^{-11} \un{m/s^2}$ is the diffusion constant of a polymer chain in the solvant. As the polymers cannot rearrange by diffusion at the interface in such a short time scale (around $1 \un{ms}$), any small surface tension gradient will be maintained and the no-slip condition justified.
Secondly, we neglect the deformability of the droplet at leading order, as already explained.
Thirdly, the hard sphere is fixed in the simulations, whereas we need to measure the stress on a moving droplet. Consequently, the medium velocity at the hard-sphere surface ($\Sigma$) has to match the experimental droplet velocity, which imposes:
\eqv{\und{u(\Sigma)}=\dd{x}{t} \cdot \und{e}_x}
where $\und{e}_x$ is the unit vector along the $x$-axis.

\paragraph*{Analysis of the simulation data.}
The analysis of the raw simulation data is performed with Matlab. Simulations provide as an output the velocity gradient for every point $M$:
\eqv{\undd{\mathrm{grad}}\,\und{u}= \left[ \begin{array}{c c c}
		u_{xx} & u_{xy} & u_{xz} \\
		u_{yx} & u_{yy} & u_{yz} \\
		u_{zx} & u_{zy} & u_{zz} \\
	\end{array} \right]}
where $u_{xy}$ is the derivative of the $x$-component of the velocity field $\und{u}$ with respect to the $y$-coordinate. For every location $x$ of the hard-sphere center, a boundary layer around the hard sphere is defined, with a thickness equal to $5 \%$ of the sphere diameter. All the analysis is performed in this boundary layer and we checked that a variation of this layer thickness does not influence our results. 
The deformation-rate tensor $\undd{d}$ is defined as the symmetric component of the previous velocity gradient :
\eqp{\undd{d}= \dfrac{1}{2} \left( \undd{\mathrm{grad}}\,\und{u} + (\undd{\mathrm{grad}}\,\und{u})^\top \right)}
In our specific geometry, we can reasonably assume that the stress is mostly in the $(x,y)$-plane. Consequently, we in fact only consider the associated bidimensional block $\undd{d}\ind{\, 2D}$ of the deformation-rate tensor, defined as follows:
\eqp{\undd{d}\ind{\, 2D}= \left[ \begin{array}{c c}
		u_{xx} & \frac{1}{2} \cdot ( u_{xy}+ u_{yx}) \\
		\frac{1}{2} \cdot ( u_{yx}+u_{xy}) & u_{yy} \\
	\end{array} \right]\label{BidimensionnalDeformationRate}}
Diagonalization of $\undd{d}\ind{\, 2D}$ gives two eigenvalues $e_1(M)$ and $e _2(M)$ for each point $M$. 
Both eigenvalues $e_1(M)$ and $e_2(M)$ are then averaged for every $M$ in the boundary layer. We would like to notice that in the description done by Barthes-Biesel \etal~\cite{BarthesBiesel1973}, a development of the flow in the vicinity of the droplet is considered and only the linear term is kept, where by construction the coefficient of the deformation rate tensor and hence the eigenvalues are constant. On the contrary, in our description, by diagonalizing locally, we get eigenvalues which depend slightly on the position $M$ along the deformed object. By averaging these values for every $M$ in the boundary layer, we get ride of the fluctuations and recover the previous description. Those averaged values, $\langle e_1 \rangle$ and $ \langle e_2 \rangle $, obtained for a given droplet position $x$, can finally be used in Equation (\ref{Dsteady}), in order to specify the position-dependent steady deformation $D_{\textrm{steady}}$ at stake.

\section{Results and discussion}
\subsection{Validation of the simulations}
As a preliminary, our numerical procedure to extract the position-dependent steady deformation $D_{\textrm{steady}}$ is tested for a known geometry (different from ours) of the literature. Specifically, we consider the setup of Hudson \textit{et al.}~\cite{Hudson2005,Cabral2006,Martin2009a}. In their geometry, the transition between the narrow channel and the wide chamber is much smoother than ours. In addition, their droplet diameter is smaller than the width (and height) of the narrow channel. Note also that, all along this test, we use a modified viscosity ratio that actually matches the one of our experiments ($\lambda=30$). As will be clear below, this specific choice enables the obtention of a second piece of information from our test (besides checking our numerical procedure): it allows to demonstrate that Hudson \textit{et al.}'s simplified approach is not valid for our experiments. 

The position-dependent steady deformation calculated by Hudson \textit{et al.}~\cite{Hudson2005} is based on Taylor's Equation~\eqref{Dsteady-Taylor} together with the following assumption:
\begin{equation}
\label{tracers}
\dot{\epsilon}= \dd{v\ind{droplet}}{x}\ ,
\end{equation}
where $v\ind{droplet}$ is the droplet velocity along the $x$-axis. Indeed, these authors assume that the droplet only acts as a tracer, \textit{i.e.} it has the exact same speed as the surrounding fluid and does not modify the flow. This assumption is motivated by both the small size of their droplets compared to the length scales of their microfluidic setup, and their low viscosity ratio ($\lambda<1$). 

We perform two simulations for Hudson \textit{et al.}'s geometry: one for the case of a tracer, \textit{i.e.} without any actual droplet (in practice we calculate the stress on a \textit{phantom} sphere), and one with a droplet (in practice we calculate the stress on a hard sphere with no slip, see~\ref{simu}). 
\begin{figure}[t!]
	\begin{subfigure}[b!]{0.44 \textwidth}
		\includegraphics[width=\textwidth]{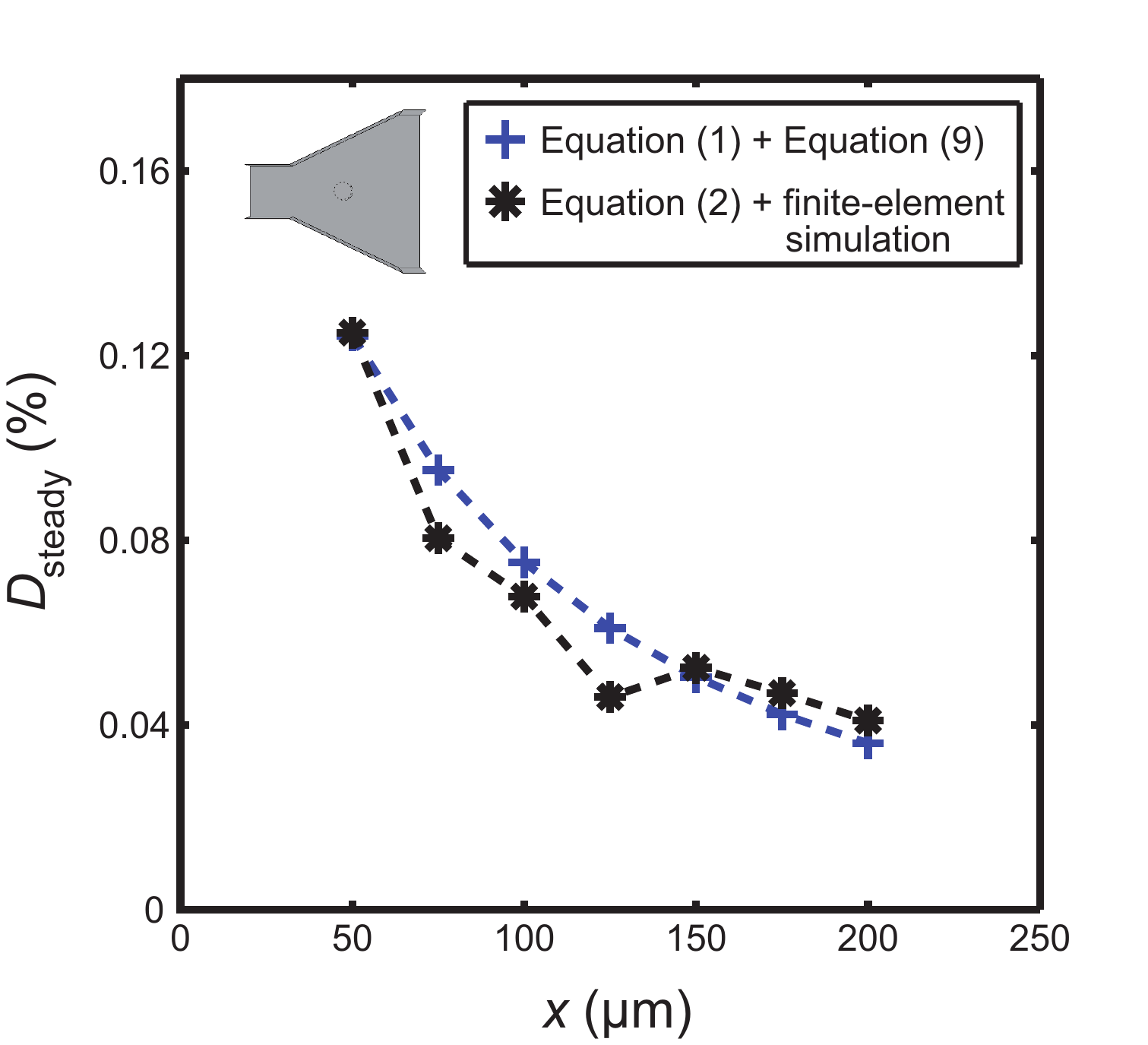}
		\caption{\label{figure2a}}
	\end{subfigure}
	\begin{subfigure}[b!]{0.44 \textwidth}
		\includegraphics[width=\textwidth]{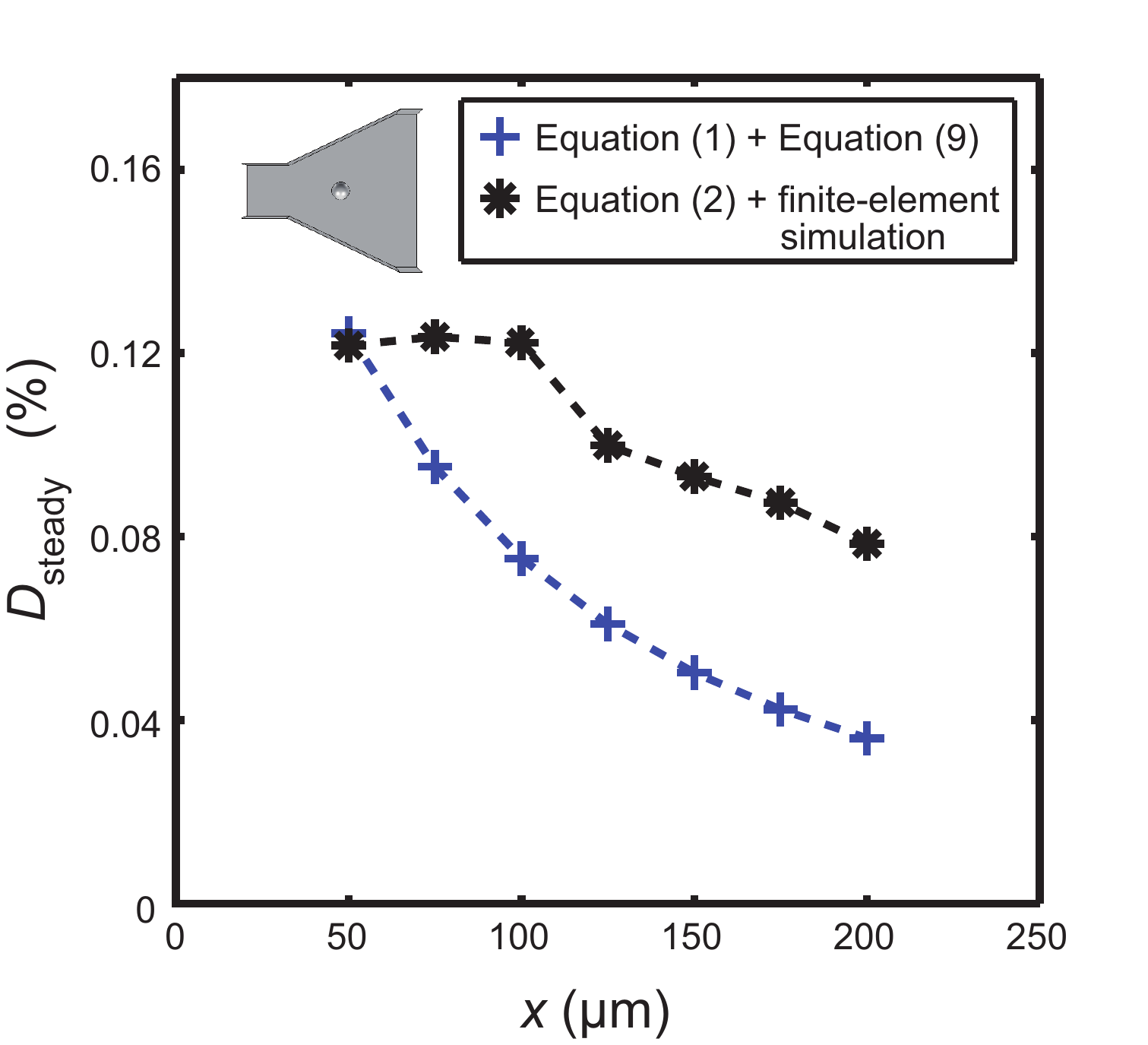}
		\caption{\label{figure2b}}
	\end{subfigure}
	\label{figure2}
	\caption{\label{figure2}(a) Steady deformation (see Equation (\ref{Dsteady})) as a function of position, as obtained from a finite-element simulation (see~\ref{simu}) of an experiment in Hudson \etal's geometry~\cite{Hudson2005} (modified viscosity ratio $\lambda=30$). The droplet is replaced by a \emph{phantom} sphere that does not disturb the flow. For comparison, we plot Equation~\eqref{Dsteady-Taylor} using Equation~\eqref{tracers}. (b) Same as (a), but when the droplet is actually modelled as a no-slip hard-sphere obstacle. For comparison, we plot Equation~\eqref{Dsteady-Taylor} using Equation~\eqref{tracers}.}
\end{figure} 
Figure~\ref{figure2} represents the comparison between the result of each of these two simulations and the model proposed by Hudson \textit{et al.} In the tracer case (Figure~\ref{figure2a}), there is a good agreement between our simulation and the model proposed by Hudson~\textit{et al.}, without any adjustable parameter. This self-consistency check indicates that our procedure calculates correctly $D_{\textrm{steady}}$ for a tracer. In contrast, in the no-slip hard-sphere case (Figure~\ref{figure2b}) there is a clear discrepancy between our simulation and the model proposed by Hudson \textit{et al.} Since we are confident in the mathematical validity of our code thanks to the self-consistency check above, we conclude that even in a non-confined and smooth geometry such as the one of Hudson~\textit{et al.}, a very viscous droplet ($\lambda=30$ here) actually modifies largely the flow field around. The model proposed by Hudson~\textit{et al.} does not apply in our case, and one therefore needs to perform a numerical simulation to know precisely $D\ind{steady}$. It is interesting to note that Taylor~\cite{Taylor1934} also observed a discrepancy between his experiments and theoretical predictions when the viscosity ratio $\lambda$ became significantly higher than $1$. 
\begin{figure}[t!]
	\begin{subfigure}[b!]{0.47 \textwidth}
		\includegraphics[width=\textwidth]{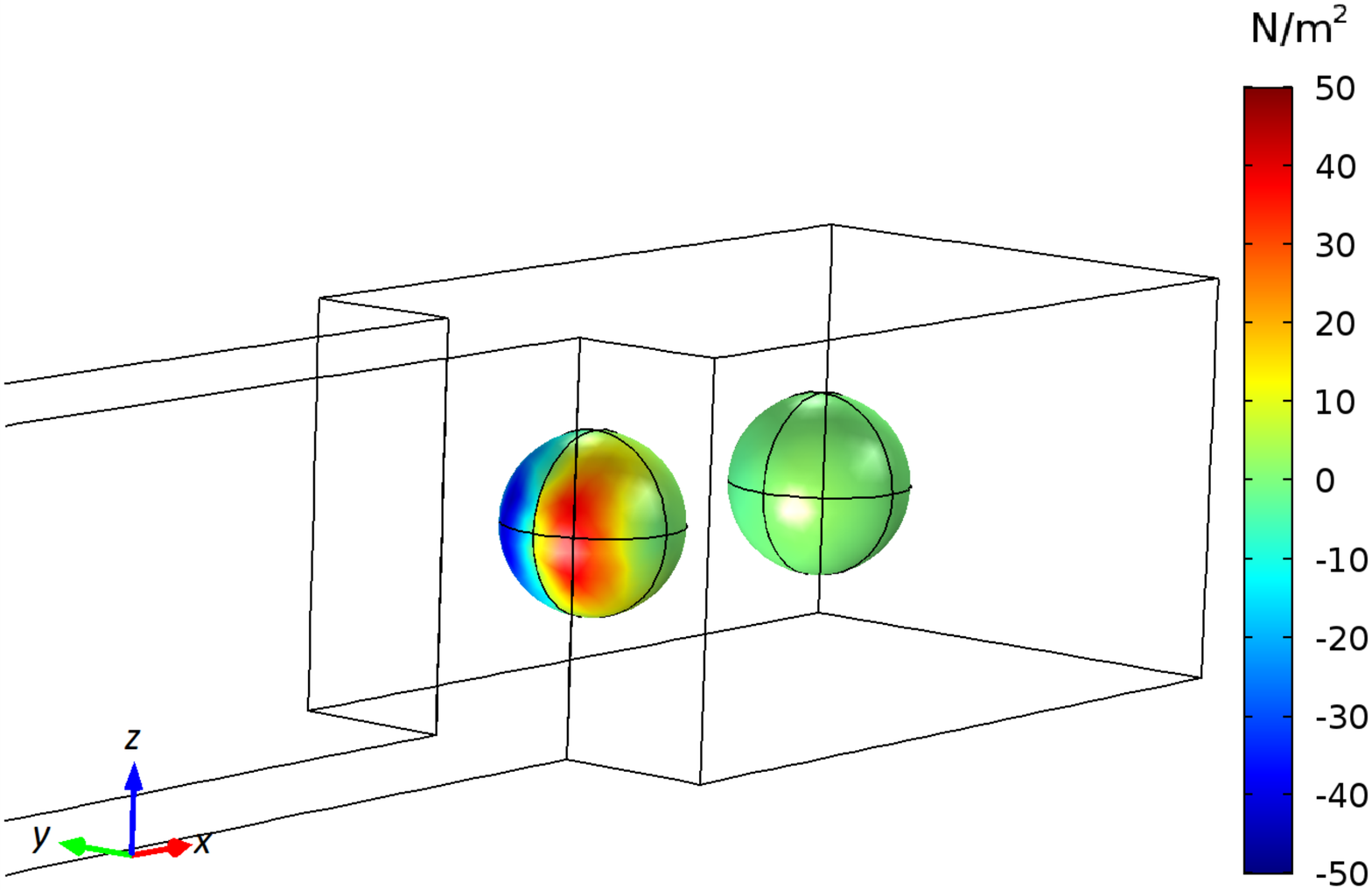}
		\caption{\label{figure3a}}
	\end{subfigure}
\\
	\begin{subfigure}[b!]{0.5 \textwidth}
		\includegraphics[width=\textwidth]{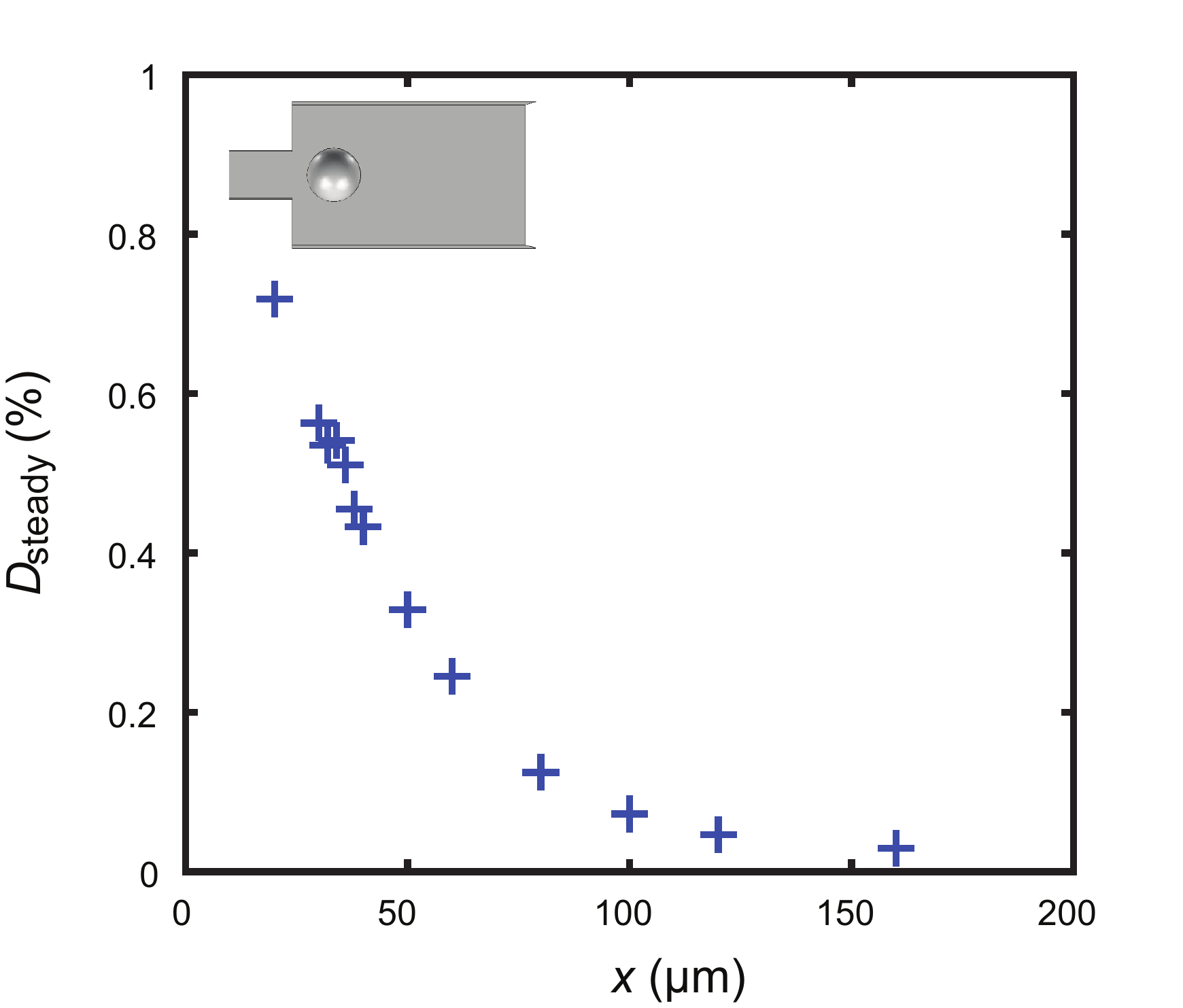}
		\caption{\label{figure3b}}
	\end{subfigure}
	\caption{\label{figure3}(a) $y$-component of the steady viscous stress at the surface of the droplet, at both the entrance and in the middle of the wide chamber, as obtained from finite-element simulations (see~\ref{simu}) of an experiment in our geometry (viscosity ratio $\lambda=30$). The droplet is modeled as a no-slip hard-sphere obstacle. (b) Resulting steady deformation (see Equation (\ref{Dsteady})) as a function of position. The steady deformation is presented here for a surface tension chosen equal to $72 \mnm$.}
\end{figure}

\subsection{Solving the governing equation} 
As explained in~\ref{simu}, the finite-element simulations allow to compute the deformation-rate tensor at the surface of the droplet as a function of the droplet position $x$ in the wide chamber, and thus the viscous stress it experiences along its trajectory (see Figure~\ref{figure3a}). From the diagonalization of the deformation-rate tensor, one can precisely calculate $D\ind{steady}(x)$ using Equation (\ref{Dsteady}) (see Figure~\ref{figure3b}). Finally, integrating Equation~\eqref{TransientD} along the trajectory of a droplet advected by the flow with the first measured point as an initial condition gives the position-dependent theoretical deformation $D\ind{th}(x)$ for a given interfacial tension.  

\begin{figure}[t!]
	\begin{subfigure}[b!]{0.45 \textwidth}
		\includegraphics[width=\textwidth]{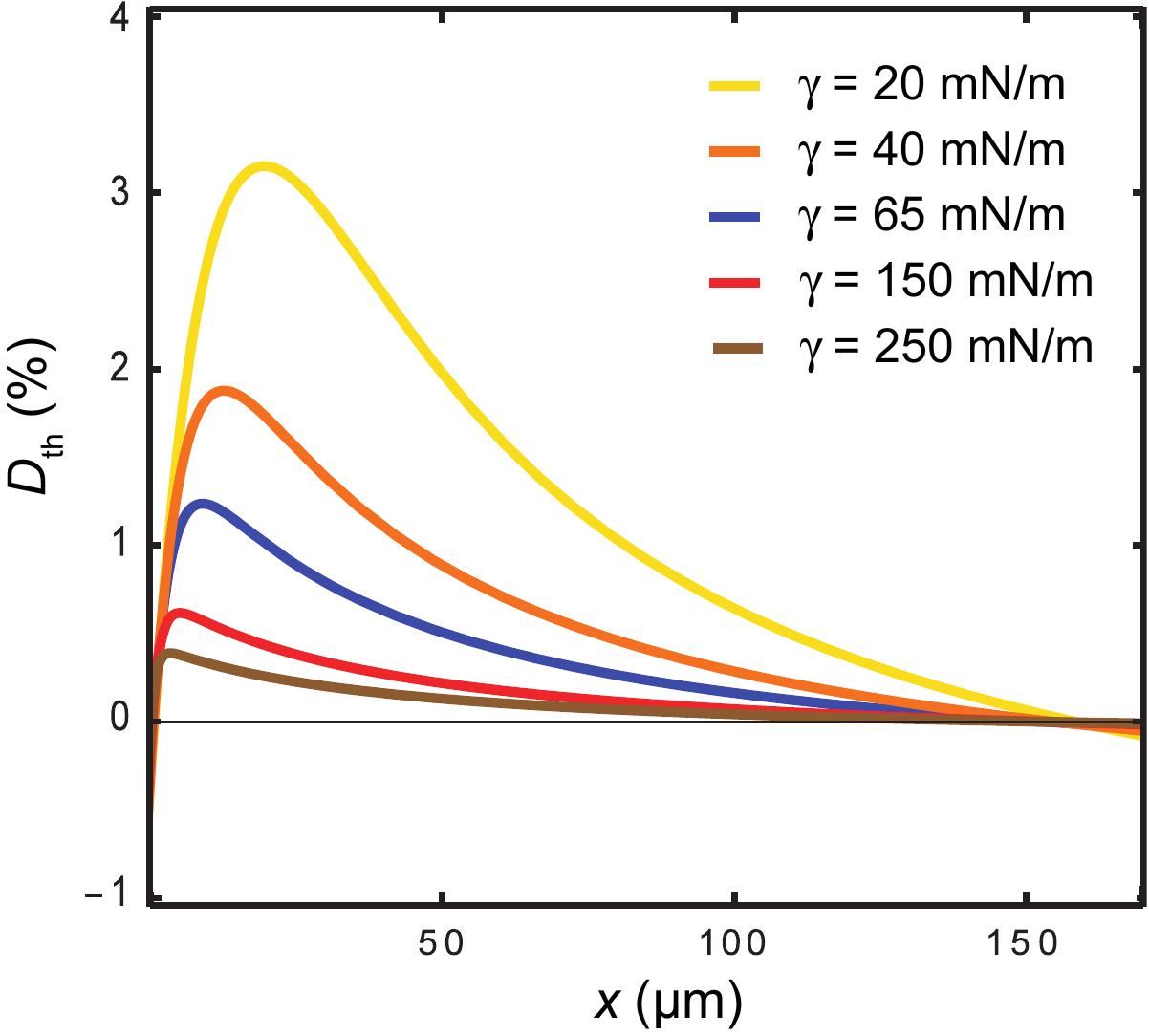}
		\caption{\label{figure4a}}
	\end{subfigure}
	\hfill
	\begin{subfigure}[b!]{0.45 \textwidth}
		\includegraphics[width=\textwidth]{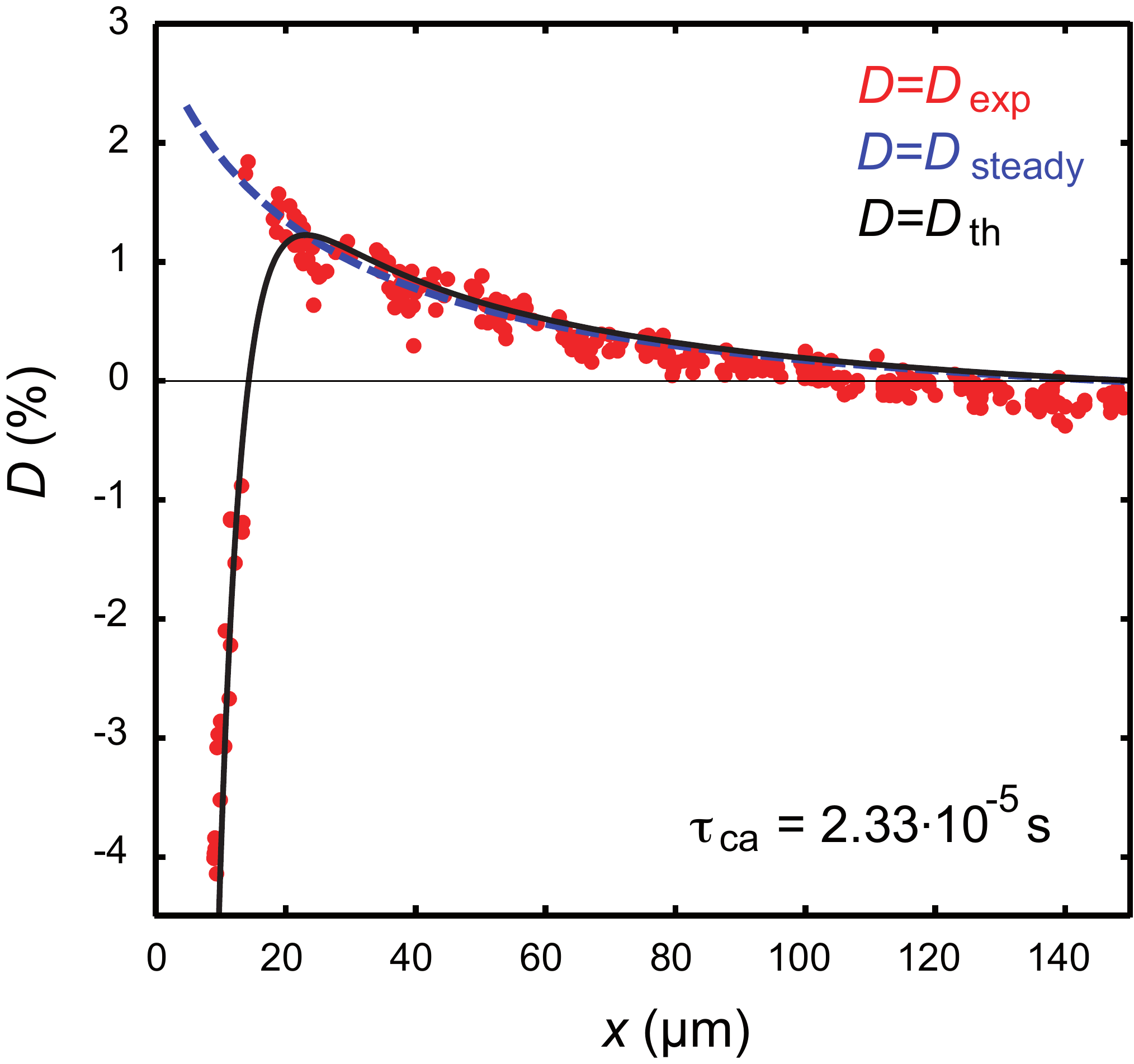}
		\caption{\label{figure4b}}
	\end{subfigure}
	\caption{\label{figure4}(a) Calculated theoretical deformation $D_{\textrm{th}}$ as a function of droplet position $x$, for various values of the droplet-medium interfacial tension $\gamma$ as indicated and an arbitrary initial condition for illustration. $D_{\textrm{th}}(x)$ is obtained by integrating Equation~\eqref{TransientD} along the trajectory of a droplet advected by the flow, and invoking the simulated $D_{\textrm{steady}}(x)$ (\fig{figure3b}). (b) Fit of the experimental data $D_{\textrm{exp}}(x)$ (dots, collected for 28 droplets) to the calculated theoretical deformation $D_{\textrm{th}}(x)$, integrated with the first data point as initial condition. The best-fit parameter $\tau_{\textrm{ca}}$ (see Equations~(\ref{TransientD}) and~(\ref{LiquidRelaxationTimeCapsule})) is indicated.}
\end{figure}

Figure~\ref{figure4a} presents different predictions of $D\ind{th}(x)$, for the droplet-medium interfacial tension $\gamma$ ranging from $20 \mnm$ to $250 \mnm$. In all cases, the deformation is initially negative, then it rapidly becomes positive and reaches a maximum value, before slowly relaxing towards zero. We also observe that an increase in interfacial tension leads to a smaller value of the maximum deformation, and to a faster overall dynamics. Therefore, thanks to the large influence of the interfacial tension on the droplet deformation in the transient regime, fitting the experimental data with those predictions enables a precise measurement of the interfacial tension. 

\subsection{Comparison with experimental data}
As it is not possible to produce stable microdroplets without surfactant, we stabilize them with poly(methacrylic acid) (PMAA)~\cite{LeTirilly2016}. In contrast to common surfactant molecules, these polymer chains do not easily desorb and can stabilize droplets, even with a relatively low quantity of molecules adsorbed. In addition, the fluctuations of the interfacial tension due to  local compression/dilatation of the surface during droplet deformation remain very low with these polymer chains. This is consistent with our model that assumes a homogeneous and constant droplet-medium interfacial tension during deformation. We note that, for more complex surface compositions, involving \textit{e.g.} surface viscoelasticity, Equation \eqref{LiquidRelaxationTimeCapsule} should be modified. This will be the object of another study.

The fitting of the experimental data, for 28 different droplets, to the theoretical predictions is shown in \fig{figure4b}. The agreement is good, and the best-fit capillary time (see Equations~(\ref{TransientD}) and~(\ref{LiquidRelaxationTimeCapsule})) is found to be $\tau\ind{ca}=2.33\pd{-5}\un{s}$. From this value, we deduce a droplet-medium interfacial tension $\gamma=35 \mypm 3 \mnm$, which is in good agreement with what can be measured using a pendant-drop apparatus ($\gamma=40 \mypm 1 \mnm$). Note that the droplet history, size and viscous-shear conditions are very different between the two experiments, which could explain the small difference. 

\section{Conclusion}
The transient deformation of a droplet in the extensional flow following a microfluidic constriction is a model problem connected to important applications, from encapsulated-drug delivery to cancer-cell detection~\cite{Byun2013}. Due to the absence of any symmetry in the associated flow and to the influence of the droplet itself on the latter, the simple approaches of the literature are not valid in general. Using finite-element simulations, we have shown that it is possible to predict the shear stress applied on the droplet and thus the resulting droplet deformation. Our model captures well our experimental data, performed with a large collection of polymer-stabilized oil microdroplets in water, which enables the robust statistical measurement of the droplet-medium interfacial tension. This model study opens the way towards precise \textit{in-situ} microrheology of capsules and cells, with more complex viscoelastic behaviours.\\

\subsubsection*{Acknowledgements}

This work was financially supported by ANR JCJC INTERPOL. The authors thank Mamisoa Nomena and Samuel Poincloux for their precious help and advice in this work. They also thank Oliver B\"aumchen, Ingmar Polenz, Julien Dupr\'e de Baubigny, Nad\`ege Pantoustier and Patrick Perrin for fruitful discussions. Finally, they thank the Global Station for Soft Matter, a project of Global Institution for Collaborative Research and Education at Hokkaido University.

\bibliographystyle{unsrt}
\bibliography{biblio}
\section*{Appendix}
\subsection*{Image processing}
A Matlab program enables to process the large number of pictures. For each of them, the background is subtracted using a reference picture with no droplet, and a threshold is automatically set to detect the droplet. For every picture containing at least one entire droplet, the position, the mean radius, and the deformation of the droplet are calculated. The droplet position $(x,\,y)$ is the mean position of all the pixels forming the droplet. 

The mean radius $r$, defined as the radius of the droplet at rest, is calculated with the two semi-axes $a$ and $b$, in the directions $x$ and $y$ respectively (see~\fig{figure1c}). For a small deformation, any deviation from the spherical shape is indeed an ellipsoid at second order in deformation, as noted by R. Cox~\cite{Cox1969} and J. Rallison~\cite{Rallison1984}. Moreover, due to the flow orientation, the axes of the droplet are along $x$ and $y$. By volume conservation, and noting $c$ the semi-axis in the $z$-direction, one gets:
\eqp{\dfrac{4}{3}\pi\cdot r^3=\dfrac{4}{3}\pi\cdot a b c\label{VolumeEllipsoid}}
We assume that when the shear flow elongates the capsule in one direction, the $z$-direction is compressed by the same amount as the other compressed direction, \textit{i.e}. $c$ is equal to the smallest semi-axis among $a$ and $b$. Consequently, using Equation~\eqref{VolumeEllipsoid}, one has:
\eqp{r= \left[ ab \cdot \min(a,b) \right]^{1/3}\label{MeanRadius}}

The deformation $D$ is then defined from the semi-axes $a$ and $b$, as follows:
\eqp{D=\dfrac{b-a}{b+a}\label{DefinitionDeformation}}
As suggested by Martin \etal~\cite{Martin2009a}, in the ellipsoidal approximation justified by Cox \cite{Cox1969}, it is in fact more precise to calculate the deformation through the moments of inertia along the $x$-axis and the $y$-axis, respectively $I_x$ and $I_y$:
\eqv{D=\dfrac{\sqrt{\vphantom{\big(}I_y}-\sqrt{\vphantom{\big(}I_x}}{\sqrt{\vphantom{\big(}I_y}+\sqrt{\vphantom{\big(}I_x}}\label{DeformationCalculation}}
where:
\eqv{I_x=\sum\ind{pixels \; i}(x_i-x)^2}
and:
\eqv{I_y=\sum\ind{pixels \; i}(y_i-y)^2}
where $x_i$ and $y_i$ are the pixel coordinates (see~\fig{figure1c}).
This is the method we employ in our study.

Consequently, the position and deformation of the droplet are known precisely for every picture containing a droplet. The time lapse between two consecutive pictures is $\Delta t=0.1 \un{ms}$ and, despite the high acquisition rate, we  have only approximately $6$ pictures per droplet due to the high flow rate. For every droplet $n$, the time $t_n$ is counted starting from the first picture where the droplet is recorded. Therefore, $t_n$ is always a multiple of $\Delta t$. The function $t_n(x)$ is fitted by a second-order polynomial function:
\eqp{t_n= a_n \cdot x^2 + b_n\cdot x +c_n \label{TimeFitX}}

All the droplets are finally synchronized using a common time measurement $t\ind{sync}$, defined as $t\ind{sync}=t_n-c_n$, and characterized by \linebreak \mbox{$t\ind{sync}(x=0)=0$}. The good droplet monodispersity, and the high acquisition rate of the camera allow to acquire enough data in a short period of time ($2 \un{s}$), during which the flow and the thermodynamic conditions remain constant. The measurements for different droplets can therefore be concatenated to generate one single curve $D_{\textrm{exp}}(x)$ (see~\fig{figure4b}).
\end{document}